\documentclass{article}
\usepackage[a4paper, margin=2cm]{geometry}
\usepackage{graphicx, multirow, booktabs} % Required for images
\usepackage{hyperref, comment, cite}
\usepackage{amsmath}
\usepackage{amssymb}
\usepackage{bm}
\usepackage{authblk} % Package for author affiliations
\usepackage{caption}
\usepackage{subcaption}
\usepackage[labelfont=bf,labelsep=period]{caption}
\usepackage{float}

\title{Time-Resolved EEG Decoding of Semantic Processing Reveals Altered Neural Dynamics in Depression and Suicidality}
\author[1,2,*]{\normalsize Woojae Jeong}
\author[2]{Aditya Kommineni}
\author[3]{Kleanthis Avramidis}
\author[4,8]{Colin~McDaniel}
\author[5]{Donald Berry}
\author[4,8]{Myzelle Hughes}
\author[9]{Thomas McGee}
\author[7]{Elsi Kaiser}
\author[7]{Dani Byrd}
\author[4,8]{Assal Habibi}
\author[6]{B.~Rael~Cahn}
\author[9]{Idan~A.~Blank}
\author[3,5]{Kristina Lerman}
\author[10]{Dimitrios Pantazis}
\author[2]{Sudarsana R. Kadiri}
\author[2]{Takfarinas Medani}
\author[2,3,5,7,8]{Shrikanth~Narayanan}
\author[1,2]{Richard M. Leahy}

\affil[1]{\normalsize Alfred E. Mann Department of Biomedical Engineering, University of Southern California, Los Angeles, CA, USA}
\affil[2]{\normalsize Ming Hsieh Department of Electrical and Computer Engineering, University of Southern California, Los Angeles, CA, USA}
\affil[3]{\normalsize Thomas Lord Department of Computer Science, University of Southern California, Los Angeles, CA, USA}
\affil[4]{\normalsize Brain and Creativity Institute, University of Southern California, Los Angeles, CA, USA}
\affil[5]{\normalsize Information Science Institute, University of Southern California, Marina Del Rey, CA, USA}
\affil[6]{\normalsize Department of Psychiatry and Behavioral Sciences, University of Southern California, Los Angeles, CA, USA}
\affil[7]{\normalsize Department of Linguistics, University of Southern California, Los Angeles, CA, USA}
\affil[8]{\normalsize Department of Psychology, University of Southern California, Los Angeles, CA, USA}
\affil[9]{\normalsize Department of Psychology, University of California, Los Angeles, Los Angeles, CA, USA}
\affil[10]{\normalsize McGovern Institute for Brain Research, Massachusetts Institute of Technology, Cambridge, MA, USA}
\affil[*]{\normalsize \textit{Corresponding author, woojaeje@usc.edu}}

\begin{document}
\maketitle

\begin{abstract}

Depression and suicidality affect cognitive and emotional processes, yet objective, task-evoked neural readouts of mental health remain limited. We investigated the spatiotemporal dynamics of affective semantic processing using multivariate decoding of time-resolved, 64-channel electroencephalography (EEG). Participants (N=137) performed a sentence-evaluation task with emotionally salient, self-referential statements. We identified robust neural signatures of semantic processing, with peak decoding accuracy between 300–600\,ms -- a window associated with rapid, stimulus-driven semantic evaluation and conflict monitoring. Relative to healthy controls, individuals with depression and suicidal ideation showed earlier onset, longer duration, and greater amplitude decoding responses, along with broader cross-temporal generalization and enhanced contributions from frontocentral and parietotemporal components. These findings suggest altered sensitivity and impaired disengagement from emotionally salient content in the clinical groups, advancing our understanding of the neurocognitive basis of mental health and establishing a compact and interpretable EEG-based index of semantic-evaluation dynamics with potential diagnostic relevance.

\end{abstract}
\section*{Introduction}

Depression is a globally prevalent mental health disorder, affecting over 280 million people, according to the World Health Organization~\cite{IHME_GBD_2023}. It is a leading cause of disability, significantly impairing cognitive, emotional, and social functioning across the lifespan~\cite{kroenke2001phq}. Depression is also a major risk factor for suicidality, including suicidal ideation, planning, and attempt. Suicidal ideation frequently co-occurs with depression and contributes to over 700,000 deaths each year~\cite{WHO_suicide}. These alarming figures emphasize the need for timely and objective methods to quantify depression and suicide risk. 

Although emerging neurophysiological methods have shown promise for the objective assessment of depression and suicidality, the field still lacks well-established neural signatures~\cite{abi2023candidate, simmatis_technical_2023}. Unlike physical conditions like heart attacks or strokes, which are characterized by well-defined physiological signatures and temporally precise indicators that guide diagnosis and intervention, mental health conditions often lack such temporally precise or functionally grounded indices. Current clinical assessments largely rely on self-report measures such as the Patient Health Questionnaire-9 (PHQ-9) and the Suicidal Ideation Scale (SIS), as well as behavioral tests like the Death Biased Implicit Association Test (D-BIAT)~\cite{kroenke2001phq, rudert2019dbiat, rudd1989prevalence}. While these tools are effective for patient screening, they may be limited as responses can be influenced by self-biases such as strategies, social desirability, or task demands~\cite{edition2013diagnostic, lieberman2007social, nisbett1977telling}, leading to underpowered assessments and clinical decisions, ultimately hindering effective intervention. Consequently, there is a critical need for objective and interpretable neural readouts that are not influenced by self-biases and that can support more accurate and robust evaluation of mental health status.

To discover reliable and objective neural readouts of depression and suicidality, it is essential to understand the underlying neural mechanisms, particularly those that occur in the early semantic processing period, i.e., prior to explicit self-reflection or behavioral response. Advances in computational neuroscience and neuroimaging have revealed widespread structural and functional abnormalities in depression, including altered activity and connectivity in the prefrontal cortex, anterior cingulate cortex (ACC), and default mode network (DMN) and salience networks~\cite{abi2023candidate, wang2012systematic, gallo2023functional}. These disruptions have been consistently observed across modalities; resting-state functional MRI (fMRI) studies report abnormal intrinsic connectivity patterns among large-scale brain networks, while electroencephalography (EEG) studies highlight atypical oscillatory dynamics during both resting and task-engaged states. At rest, remitted depression shows elevated EEG functional connectivity between posterior cingulate (PCC) and subgenual prefrontal regions (sgPFC)~\cite{benschop_electrophysiological_2021}. During tasks, major depression exhibits reduced task-induced upper-alpha desynchronization during working memory, consistent with abnormal inhibition~\cite{segrave2010alpha}, with additional oscillatory ERP signatures (altered frontal alpha asymmetry, increased theta, reduced P300 amplitude, and longer latency) synthesized in recent reviews~\cite{simmatis_technical_2023, deaguiar2019review}. Notably, semantic processing has emerged as particularly sensitive to depressive symptomatology. Event-related potential (ERP) studies have identified reduced N400 amplitude and attenuated late positive potentials in depressed individuals during evaluation of emotionally salient or self-relevant contents~\cite{iakimova2009behavioral, kiang2017abnormal, shestyuk2005reduced, shestyuk2010representation, auerbach2015self}. These disruptions suggest impairments in integrative and evaluative processing that are not captured by the resting-state paradigms alone. 

Despite these advances, research comparing neural responses to affective semantic stimuli across control, depressed, and suicidal populations remains scarce, particularly in paradigms that closely mimic the demands of clinical self-report questionnaires. Such tasks require participants to actively evaluate self-referential and emotionally charged statements, thereby more closely resembling clinical evaluations than passive or resting-state paradigms, and may better capture clinically meaningful neural dynamics. Moreover, to fully understand the underlying neurocognitive mechanisms, it is essential to examine both the spatial and temporal dynamics of brain activity, as neural processing is inherently distributed across regions and evolves rapidly over time. However, conventional neuroimaging methods have limitations. Functional MRI offers high spatial resolution but lacks the temporal fidelity necessary to capture sub-second neural activity. In contrast, traditional ERP approaches, while temporally precise, often rely on univariate analyses centered on a small number of electrodes or components, limiting the ability to resolve distributed spatial patterns~\cite{luck2014erp, cichy_multivariate_2017}. These constraints may obscure key features of neural computations that emerge from the coordinated activity of brain networks over time.

Here, we take an EEG-based approach that targets the spatiotemporal dynamics of semantic evaluation during a sentence evaluation task that actively engages subjects in responding to emotionally salient, self-referential statements~\cite{hughes2025precog}. Multivariate pattern analysis has been widely used to investigate the temporal dynamics of neural activity and to assess whether distinct neural patterns emerge in response to varying cognitive or emotive conditions~\cite{dobs_how_2019, jeong_multivariate_2023, jeong_motion_2019, wolff_revealing_2015, wolff_dynamic_2017, myers_testing_2015}. By combining this task design with multivariate EEG decoding, we aimed to capture the spatiotemporal dynamics of semantic processing, enabling a more holistic view of how neural representations evolve across mental health conditions. We first examined whether sentences differing in semantic meaning elicited distinguishable neural representations across all participants, focusing on the early stage of semantic processing that is minimally influenced by participants' active decision making or behavioral response. We then assessed whether these neural signatures systematically varied across groups with different mental health status.

Our results revealed that sentences with different semantic meanings elicited distinct and temporally dynamic neural patterns, with peak differentiation emerging during the early semantic processing window (300-600 ms), a period typically associated with stimulus-driven semantic evaluation, conflict monitoring, and integrative processing~\cite{brouwer2013time, friederici2011brain, kutas1980reading, troyer2020catch,  lau_cortical_2008}. This effect was primarily driven by frontocentral and parietotemporal components associated with semantic evaluation and integration~\cite{brouwer2013time, friederici2011brain, troyer2020catch, lau_cortical_2008, kutas2011thirty, zhou_posterior_2019, binder2009semantic, valdebenito-oyarzo_parietal_2024}. Individuals with depression or suicidality exhibited earlier, stronger, and more sustained decoding responses than controls, reflecting heightened sensitivity and diminished disengagement from emotionally charged content. Taken together, these time-resolved EEG signatures provide a compact and interpretable index of semantic evaluation dynamics and modulation of mental state, potentially informing future diagnostic development pending prospective validation.
\section*{Results}

We analyzed behavioral responses and EEG recordings from 137 participants: 48 healthy controls (C), 41 individuals with depression without suicidality (D), and 48 individuals with depression and suicidality (S), who completed a sentence evaluation task (Fig.~\ref{fig1}A, see Methods for details). First, we analyzed the behavioral responses to determine whether participants reacted differently to sentences based on their semantic content. Then, we conducted within-subject EEG decoding to examine the spatiotemporal dynamics of neural representations associated with semantic processing. By comparing these neural responses across groups, we identified differences associated with mental health status. Finally, we analyzed the underlying spatiotemporal characteristics to understand the sources of these group-level differences. 

\subsection*{Response time analysis during the sentence evaluation task}

Participants completed 320 trials in which each sentence was presented word by word. After the final word of each sentence, participants were asked to indicate whether they agreed or disagreed with the sentence. Congruency labels were defined based on the expected response of group S (see methods for details). In this context, the terms congruent and incongruent are used purely as predefined categorical labels rather than reflecting group-specific semantic evaluations. That is, while the subjective interpretation of a sentence may differ across groups, here, congruency simply denotes the label assigned according to the group S reference, which was applied consistently across all groups for analyses. In general, congruent sentences carried negative semantic content, while incongruent sentences carried neutral or positive content (Table S1).
%(Table~\ref{tabs1}).

\begin{figure}[H]
    \centering
    \includegraphics[width=0.9\textwidth]{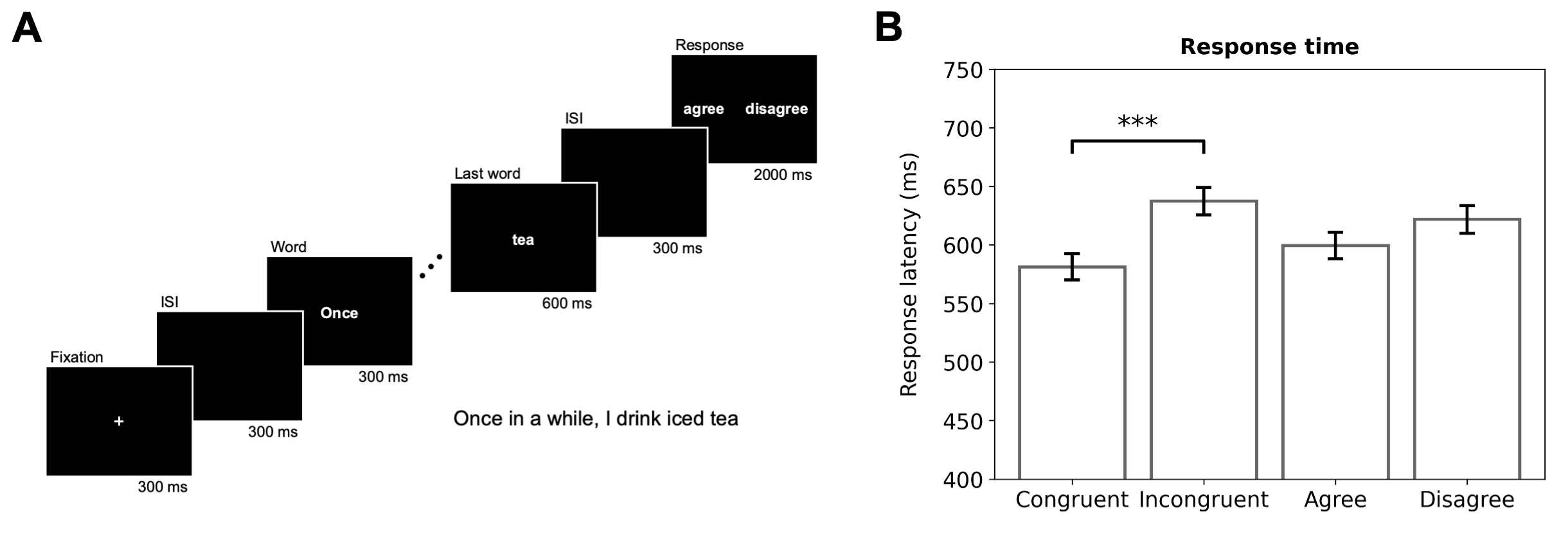}
    \caption{\textbf{Sentence evaluation task and behavioral responses.}
    \textbf{A.} Participants were presented with 320 sentences and prompted to indicate whether they agreed or disagreed with each sentence. Each trial began with a 300\,ms fixation cross, followed by sequential presentation of each word in the sentence for 300\,ms, with a 300\,ms black screen inter-stimulus-interval (ISI) between words, except for the final word, which was presented for 600\,ms. Participants then had a 2-second window to respond by indicating “agree” or “disagree” using a push-button box.
    \textbf{B.} Average response time across participants for different sentence types (congruent and incongruent) and responses (agree and disagree). Responses that were faster than 100\,ms were discarded. The error bars denote standard error. Stars correspond to statistically significant differences between conditions (Welch's \textit{t}-test, $***$ \(p<0.001\)).}
    \label{fig1}
\end{figure}

To examine whether individuals, irrespective of their mental health status, interpret semantically charged sentences differently, we analyzed their behavioral responses. Mean response times were computed to assess the cognitive effort associated with interpreting the sentences (Fig.~\ref{fig1}B). Participants responded significantly faster to congruent sentences compared to incongruent ones (\(581.18\pm11.23~ms\) vs.\ \(637.33\pm11.55~ms\), Welch's \textit{t}-test, \(p=0.0006\)). This implies that congruent sentences, which predominantly conveyed negative sentiment, were processed more rapidly than incongruent sentences, which more often carried positive or neutral emotional content. Although participants did not show a statistically significant difference in response times when agreeing versus disagreeing with the sentence\textemdash regardless of its sentiment\textemdash we observed a trend toward longer response times during disagreement (\(599.74\pm11.29~ms\) vs.\ \(621.72\pm11.71~ms\), Welch's \textit{t}-test, \(p=0.1737\)).

The results demonstrate that individuals exhibit distinct behavioral patterns when assessing sentences for their semantic meaning and alignment with personal expectations~\cite{auerbach2015self}. The observed differences in response times suggest that emotionally and semantically charged content engages distinct processing mechanisms, likely reflecting its relevance to individuals' internal emotional states and contextual expectations. This behavioral variability implies that the underlying neural representations during semantic processing may also diverge across conditions. To explore this further, we investigated how these differences are reflected in the underlying neural mechanisms.

\subsection*{Temporal dynamics of semantic processing from multivariate EEG decoding}

\begin{figure*}[t]
    \centering
    \includegraphics[width=\textwidth]{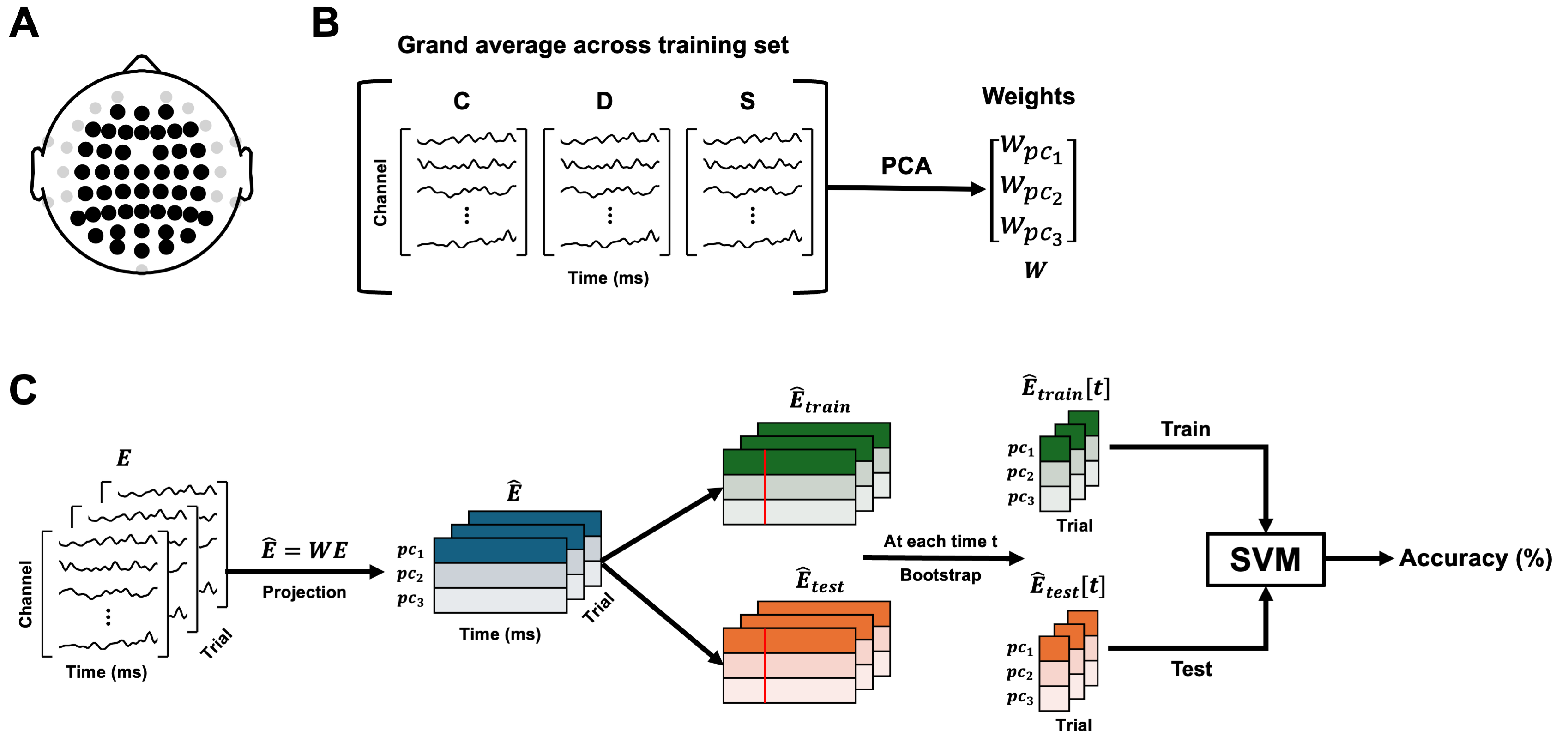}
    \caption{\textbf{Multivariate EEG pattern analysis pipeline}
    \textbf{A.} Channel topography of  64-channel configuration. All peripheral channels were excluded from analysis to avoid noisy inputs. Black dots indicate the inner set of 47 channels used in the EEG analysis.
    \textbf{B.} Latent features shared across groups (C: control, D: depressed, and S: suicidal) were extracted using PCA on the grand-averaged EEG data across the training set (see Methods for details). Every trial in the training set of all participants within a group was averaged to obtain the grand-averaged EEG per group. The resulting grand-averaged EEGs were concatenated across time, and PCA was performed. The weight vectors for the top three PCs $\left(\bm{w}_{\bm{pc}_{1}},\ \bm{w}_{\bm{pc}_{2}},\ \bm{w}_{\bm{pc}_{3}}\right)$ were retained, forming the projection weight matrix $\bm{W}$. \textbf{C.} Within-subject EEG decoding using linear kernel SVM was applied. Each subject's EEG data $\bm{E}$ was projected onto a shared latent space using $\bm{W}$, yielding the latent representation of the neural data $\bm{\hat{\bm{E}}}$. Then, 250 bootstrapped trials were generated for each condition (200 for training and 50 for testing) by randomly sampling with replacement and sub-averaging 12 trials independently for each training/testing cohort. For each time point $t$, a 5-fold cross-validation using an SVM classifier was performed. Decodability (\%) was reported by averaging the accuracies across all folds (see Methods for details).}
    \label{fig2}
\end{figure*}

To examine differences in neural representations during the processing of sentences with different semantic meaning, we performed a within-subject, multivariate pattern analysis (Fig.~\ref{fig2}, see Methods for details) on the EEG data recorded during the task. The window of EEG analysis spanned from 200\,ms before to 1500\,ms after the onset of the final word of the sentence, capturing a pre-onset baseline period (-200-0\,ms), the stimulus presentation period (0-600\,ms) and a portion of the response period (900-1500\,ms). For each subject, trials were divided into five stratified folds. Shared latent features (Fig.~\ref{fig2}B) were derived by applying principal component analysis (PCA) to EEG data that were first averaged across training trials (four folds) within each subject, then averaged across subjects within each group, and finally concatenated across time. We constructed a projection matrix $\bm{W}$, composed of the weight vectors corresponding to the top three principal components. Then all individual EEG trials $\bm{E}$ were projected onto the shared latent space as $\bm{\hat{\bm{E}}} = \bm{WE}$. As the final stage, decoding was performed using binary classification on 500 augmented trials generated through bootstrap averaging, enabling determination of the sentence type or response type at each time point. A linear Support Vector Machine (SVM) was used for classification. The decodability measure was quantified by averaging decoding accuracies from 5-fold cross-validation within each subject, and then averaging across all 137 participants (Fig.~\ref{fig2}C). This analysis investigated whether patterns of neural activity could be distinguished within each subject when viewing different sentence types or producing different responses.

We first investigated the temporal characteristics of the semantic processing using EEG-based decoding analysis. Semantic processing unfolds in a sequence of distinct stages, including early semantic evaluation, integration, and conflict monitoring, followed by later sustained semantic analysis and decision-making~\cite{kutas1980reading, lau_cortical_2008, friederici2011brain, brouwer2013time, binder2009semantic, troyer2020catch, zhou_posterior_2019}. The earlier phase is generally considered to reflect pre-attentive neural activity that occurs prior to active decision making, while the later phase involves conscious appraisal~\cite{kutas1980reading, friederici2011brain, troyer2020catch, del2007brain}. Our primary focus was to examine neural responses to semantic content, distinct from those associated with active decision-making processes. To disentangle these effects, we conducted separate decoding analyses for sentence types and response types, allowing us to compare the neural signatures associated with semantic evaluation and those driven by behavioral responses, respectively. 

\begin{figure*}[t]
    \centering
    \includegraphics[width=\textwidth]{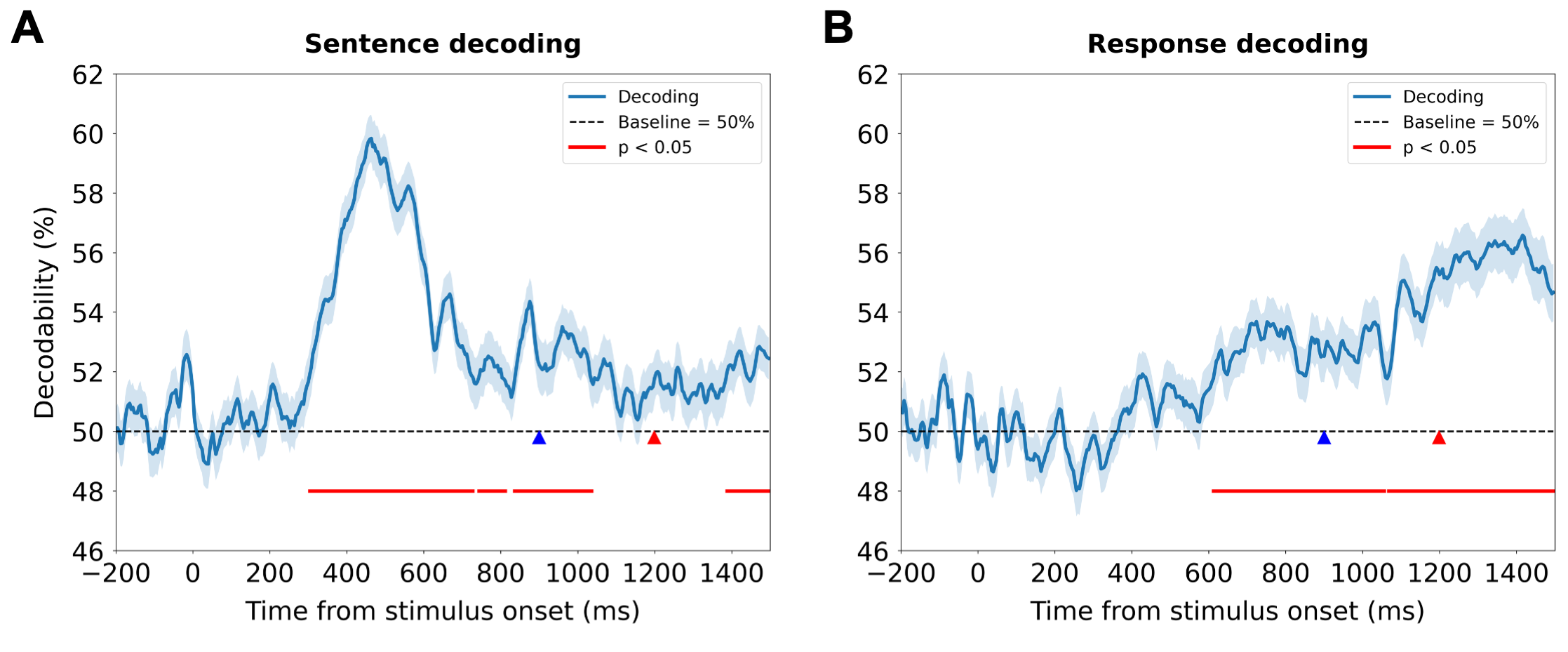}
    \caption{\textbf{EEG decoding of sentence semantics and behavioral responses}
    \textbf{A, B.} Time-resolved decoding accuracy (\%, decodability) of multivariate EEG patterns for (\textbf{A}) congruent vs.\ incongruent sentence evaluation and (\textbf{B}) agree vs.\ disagree responses, aligned to the onset of the final word of the sentence (\(N=137\), solid blue lines). Red horizontal lines along the x-axis indicate time points, where decoding accuracy was significantly above chance, as determined by a two-sided cluster-based permutation test (\(N=137\), cluster-defining threshold \(p<0.05\), corrected significance level \(p<0.05\), 5000 permutations). Shaded blue areas represent the standard error of the mean. The response window began at 900\,ms (blue triangle), and the red triangle marks the mode of response time across all participants (1199\,ms).}
    \label{fig3}
\end{figure*}

For the sentence decoding analysis, 250 bootstrapped trials (200 for training and 50 for testing) per condition (congruent and incongruent) were generated within each subject. Each bootstrapped trial was computed by averaging randomly sampled 12 trials (approximately 10\% of the trials in each congruency condition of the training set) with replacement from the same condition within either the training or testing cohort, ensuring that bootstrapping was performed independently for the training/testing dataset. Each bootstrapped trial preserved the semantic content of the respective condition type while reducing the effect of individual responses, which allows investigating neural responses related to semantic perception. This approach was also chosen to minimize confounding effects from active decision making, response preparation, and motor execution. Neural decodability between congruent and incongruent conditions was significantly above baseline starting from 304\,ms after onset, with decoding performance peaking at 464\,ms (mean accuracy: 59.83\%, Fig.~\ref{fig3}A). The earliest significant decoding cluster spanned from 304-728\,ms and persisted through additional time windows (744-812\,ms and 836-1036\,ms), as confirmed by a two-sided cluster-based permutation test, \(N=137\), \(p<0.05\)). 
A second set of significant clusters was observed near the average behavioral response time (1448.6\,ms), spanning 1388-1444\,ms and 1452-1500\,ms (two-sided cluster-based permutation test, \(N=137\), \(p<0.05\)), with lower peak decoding accuracy of 52.84\%. These results suggest that the early decoding cluster with peak decodability (304-728\,ms) likely represents initial semantic perception and information integration of stimulus meaning with prior expectations, and conflict monitoring. The later clusters (744-812\,ms and 836-1036\,ms) may correspond to sustained semantic integration processes and perceptual decision making, while the final clusters near the behavioral response window (1388-1444\,ms and 1452-1500\,ms) likely reflect motor execution.

To further support our interpretation of the temporal dynamics observed in sentence decoding, we conducted a complementary decoding analysis based on participants' behavioral decisions. Using an identical multivariate classification framework, decoding was performed to examine whether neural activity patterns could reliably distinguish between trials in which participants agree vs.\ disagree with the presented sentences. Here, the entire analysis pipeline, from training/testing split through classification, was performed separately from the sentence decoding, using trial labels re-categorized based on each subject’s behavioral response (agree or disagree). This approach allowed us to isolate neural signatures associated with the evaluative decision-making process as well as response preparation and execution. Significant response decoding emerged 612\,ms onwards, and remained significant throughout the response period (612-1056\,ms, 1068-1496\,ms, two-sided cluster-based permutation test, \(N=137\), \(p<0.05\), Fig.~\ref{fig3}B). The earlier sustained significant cluster from response decoding (612-1056\,ms) corresponds to the perceptual decision-making, consistent with prior studies~\cite{brouwer2013time, zhou_posterior_2019}. The latter response decoding cluster showed increased decodability, aligning with the mode response time (1199\,ms, Fig.~\ref{fig3}B, red triangle) with peak performance (mean accuracy: 56.68\%) occurring at 1416\,ms, closely aligning with the mean response time (1513\,ms) across participants %(Supplementary Fig.~\ref{supfig1}). 
(Supplementary Fig.\,S1). These findings support our interpretation of the temporal dynamics observed in the sentence decoding results, suggesting that the observed neural dynamics in the earlier period (300-600\,ms) likely reflect cognitive processes such as semantic perception, conflict monitoring, and integration with prior expectations that precede active decision-making~\cite{friederici2011brain}.

\begin{figure*}[t]
    \centering
    \includegraphics[width=\textwidth]{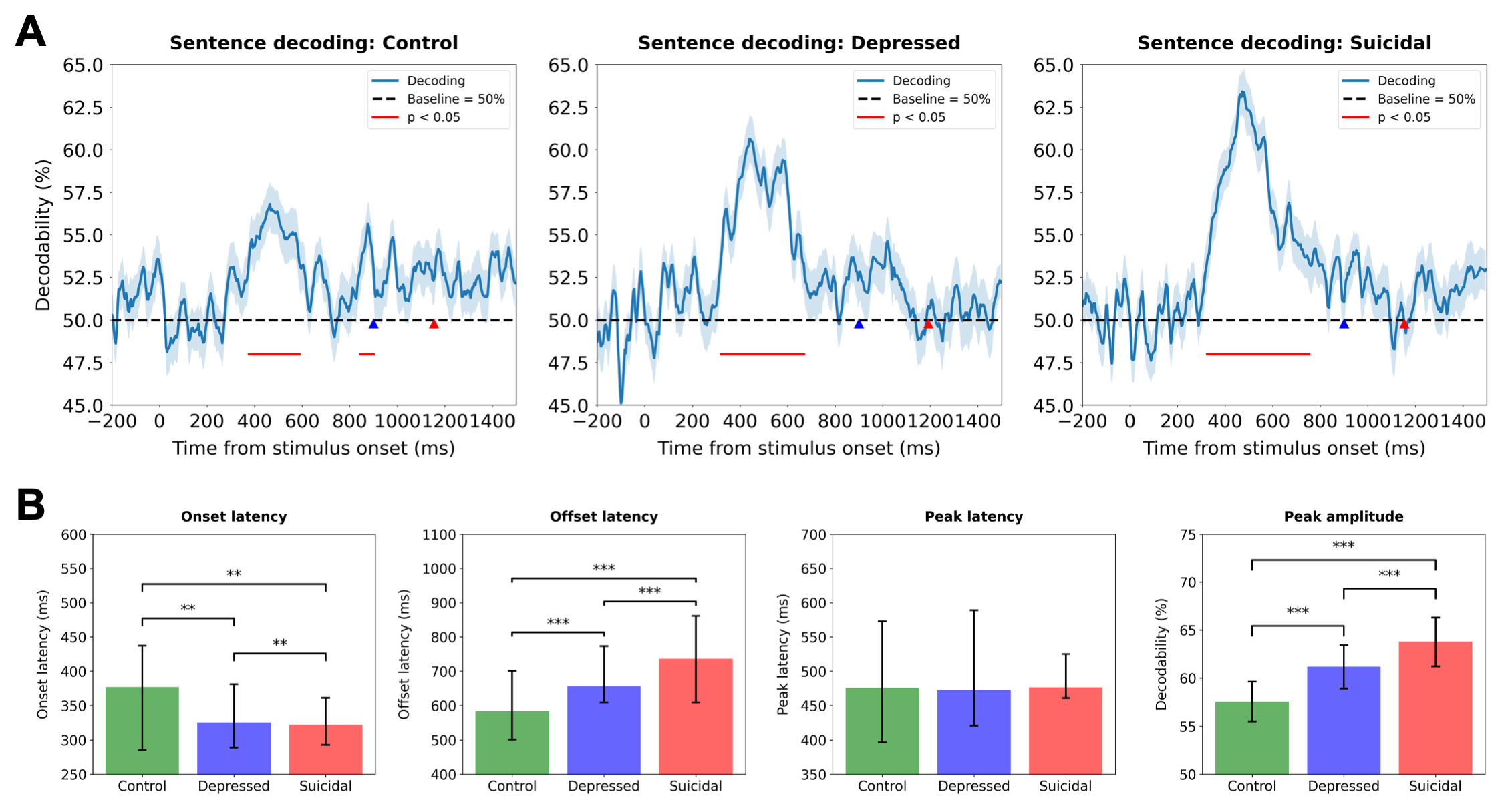}
    \caption{\textbf{Group sentence decoding from EEG}
    \textbf{A.} Sentence decoding averaged across participants within each group (control, depressed, and suicidal) aligned to the onset of the final word (solid blue lines). Red-colored lines at the bottom of each plot indicate time points where the decoding accuracy (decodability, \%) was significantly above chance level (50\%, dashed black lines; two-sided cluster-based permutation test, cluster-defining threshold \(p<0.05\), corrected significance level \(p<0.05\), 5000 permutations). Blue triangle indicates the response window start (900\,ms), and red triangle shows the mode of the response time for each group (control: 1154\,ms, depressed: 1192\,ms, and suicidal: 1154\,ms).
    \textbf{B.} Onset latency, offset latency, peak latency, and peak amplitude for sentence decoding in each group (control: green, depressed: blue, and suicidal: red). Error bars denote bootstrapped 95\% confidence intervals. Stars indicate significant difference between groups (one-sample two-sided bootstrap test, $***$ \(p<0.001\), $**$ \(p<0.01\), FDR-corrected).}
    \label{fig4}
\end{figure*}

\subsection*{Group differences in semantic decoding}

The preceding results demonstrated that different sentence types elicit significantly distinguishable neural representations. Next, we investigated whether these neural representations differ across groups with distinct mental health profiles. Participants exhibited distinct behavioral response patterns to the same set of sentences across groups %(Table~\ref{tabs2}). 
(Table S2). Individuals in the control group were more likely to disagree with congruent\textemdash typically framed with negative valence\textemdash sentences and to agree with incongruent sentences, which were generally positively or neutrally valenced. In contrast, individuals in the suicidal group exhibited the opposite trend, showing a greater tendency to agree with congruent sentences and disagree with incongruent sentences. The depressed group showed an intermediate response pattern between the control and suicidal groups. These response patterns suggest that evaluative biases toward emotionally salient content vary systematically with mental health status, and that such differences may be reflected in the underlying neural representations.

To examine group-level differences, we first analyzed group-wise sentence decoding by averaging participant-level decoding results within group (Fig.~\ref{fig4}A). All three groups showed significant above-chance decoding: control (376-588\,ms and 844-900\,ms, two-sided cluster-based permutation test, \(N=48\), \(p<0.05\)), depressed (320-668\,ms, two-sided cluster-based permutation test, \(N=41\), \(p<0.05\)), and suicidal (324-752\,ms, two-sided cluster-based permutation test, \(N=48\), \(p<0.05\)). Although the overall decoding patterns were qualitatively similar across groups, they differed in temporal onset, duration, and peak amplitude.

To quantitatively compare decoding dynamics between groups, we analyzed the onset latency, offset latency, peak latency, and peak amplitude of the decoding curves using bootstrap tests (see Methods for details). In this analysis, we focused on the characteristics of the significant decoding cluster that encompassed the peak decoding accuracy in the sentence decoding results (Fig.~\ref{fig4}A, main significant clusters). Onset and offset latencies were defined as the first and last points of the significant cluster, respectively. The onset of significant decoding in the control group arose significantly later (376.95\,ms, 95\% CI: 285--437\,ms) than in the clinical groups (depressed: 325.97\,ms, 95\% CI: 289--381\,ms and suicidal: 322.25\,ms, 95\% CI: 293--361\,ms) as tested by one-sample two-sided bootstrap test (\(p<0.001\), FDR-corrected, Fig.~\ref{fig4}B, onset latency). Although the onset latency difference between the clinical groups was statistically significant (one-sample two-sided bootstrap test, \(p<0.01\), FDR-corrected), the actual difference was minimal (3.42\,ms). The offset latency significantly differed across all groups, with the control group exhibiting the earliest offset, followed by the depressed group, and the suicidal group showing the latest offset (control: 583.99\,ms, 95\% CI: 501--701\,ms, depressed: 656.35\,ms, 95\% CI: 609--773\,ms, and suicidal: 736.37\,ms, 95\% CI: 609--861\,ms, one-sample two-sided bootstrap test, \(p<0.001\), FDR-corrected, Fig.~\ref{fig4}B, offset latency). When considered alongside the onset latency, these results indicate that the control group had the shortest sustained decoding, followed by the depressed group, while the suicidal group showed the most prolonged decoding duration (control: 207.04\,ms, 95\% CI: 80--328\,ms, depressed: 330.68\,ms, 95\% CI: 228--472\,ms, and suicidal: 414.12\,ms, 95\% CI: 264--548\,ms, one-sample two-sided bootstrap test, \(p<0.001\), FDR-corrected). 

While peak latency did not differ significantly across groups (control: 475.83\,ms, 95\% CI: 397--573\,ms, depressed: 472.46\,ms, 95\% CI: 421--589\,ms, and suicidal: 476.55\,ms, 95\% CI: 461--525\,ms, Fig.~\ref{fig4}B, peak latency), we observed significant difference in peak decoding performance across all groups (control: 57.52 \%, 95\% CI: 55.5--59.6 \%, depressed: 61.17\%, 95\% CI: 58.9--63.4 \%, and suicidal: 63.78 \%, 95\% CI: 61.2--66.3 \%, one-sample two-sided bootstrap test, \(p<0.001\), FDR-corrected, Fig.~\ref{fig4}B, peak amplitude). These results indicate that individuals with depressive or suicidal states exhibited earlier engagement, longer sustained decoding, and stronger neural discriminability of sentence meaning. This enhanced, earlier, and prolonged decoding in clinical groups may reflect higher sensitivity to emotionally charged semantic content, suggesting that the dynamics of semantic processing vary depending on the mental state.

\subsection*{Time-resolved spatial contribution of neural features}

\begin{figure*}[t]
    \centering
    \includegraphics[width=0.8\textwidth]{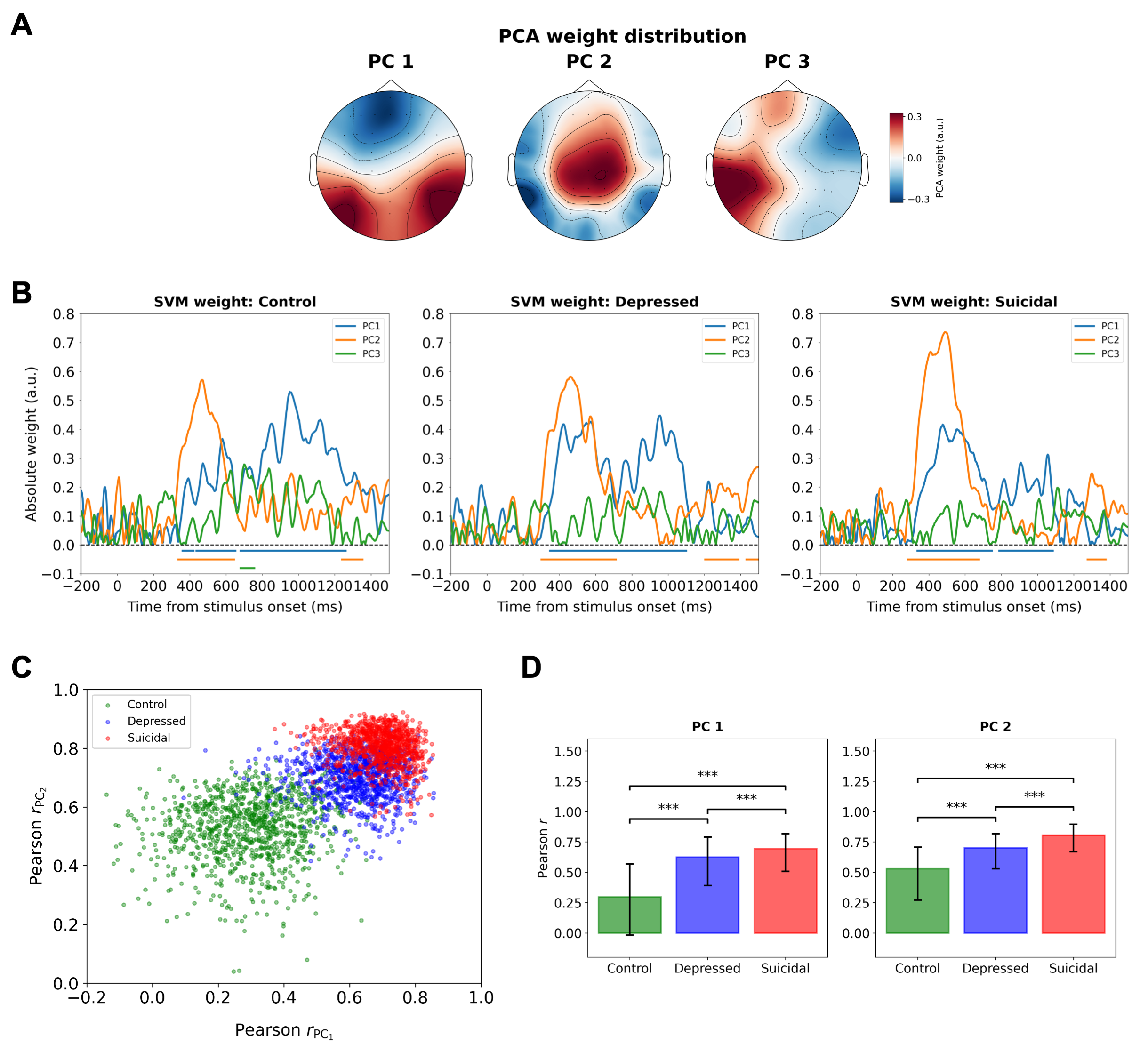}
    \caption{\textbf{Spatial contribution estimates of the sentence decoding neural features}
    \textbf{A.} Topographical distributions of the top 3 PCs estimated from the multivariate analysis (Fig.~\ref{fig2}B). Red-highlighted regions indicate the enhanced channel contribution, and the blue-highlighted regions indicate the suppressed channel contribution.
    \textbf{B.} Mean estimation of the weight assigned to each of the 3 PCs by the linear SVM across participants at each time point relative to the final word onset during sentence decoding. The colored lines at the bottom of the plot denote the significant time points (two-sided cluster-based permutation test, cluster-defining threshold \(p<0.05\), corrected significance level \(p<0.05\), 5000 permutations).
    \textbf{C, D.} Relationship between sentence decoding accuracy and spatial feature contributions across groups (control: green, depressed: blue, and suicidal: red). \textbf{C}. Empirical distribution of the Pearson correlation between sentence decoding accuracy (Fig.~\ref{fig4} A) and time-resolved absolute SVM feature weights (Fig.~\ref{fig5} B) for each group. The x-axis represents the Pearson correlation with \(PC_1\) weights, and the y-axis denotes the Pearson correlation with \(PC_2\). \textbf{D}. Group-level comparison of correlation coefficients for \(PC_1\) (left) and \(PC_2\) (right). Error bars denote bootstrapped 95\% confidence intervals. Stars indicate significant difference between groups (one-sample two-sided bootstrap test, $***$ \(p<0.001\), FDR-corrected).}
    \label{fig5}
\end{figure*}

To characterize the spatial activations underlying sentence decoding, we analyzed the weights of the top 3 PCs, estimated from the multivariate analysis (Fig.~\ref{fig2}B). The topographic maps of the estimated PC weights illustrate the spatial contribution of each EEG channel to the corresponding latent features~\cite{spencer2001spatiotemporal, tenke2008hemispatial}. We found that \(PC_1\) exhibited strong activation in the occipital and parietotemporal regions, which are typically involved in early visual processing, sustained semantic integration, and perceptual and categorical decision-making (Fig.~\ref{fig5}A, left) ~\cite{zhou_posterior_2019, binder2009semantic}. \(PC_2\) showed prominent activation in the central and anterior channels involved in conflict monitoring, semantic integration, decision formation, and response selection (Fig.~\ref{fig5}A, middle)~\cite{miller2001integrative}. \(PC_3\) displayed a left-lateralized activation, aligning with the well-established dominance of the left hemisphere in language and semantic processing (Fig.~\ref{fig5}A, right) ~\cite{friederici2011brain}. These findings suggest that PCA effectively extracted spatial features relevant to semantic perception, information integration, and the decision-making process during sentence evaluation.

Next, to examine how each feature contributed to sentence discrimination, we analyzed the time-resolved weights assigned to each of the PCs by the linear SVM classifier at each time point, separately for each group. In all groups, we found that \(PC_1\) and \(PC_2\) showed significant contribution corresponding to the period of significant sentence decodability (Fig.~\ref{fig4} A, significant clusters). \(PC_2\) exhibited the most prominent contribution during the peak decoding period: control (332-640\,ms, two-sided cluster-based permutation test, \(N=48\), \(p<0.05\)), depressed (296-708\,ms, two-sided cluster-based permutation test, \(N=41\), \(p<0.05\)) and suicidal (280-672\,ms, two-sided cluster-based permutation test, \(N=48\), \(p<0.05\), Fig.~\ref{fig5} B, orange). This was accompanied by a sustained contribution from \(PC_1\): control (356-416, 432-648\,ms, two-sided cluster-based permutation test, \(N=48\), \(p<0.05\)), depressed (344-1096\,ms, two-sided cluster-based permutation test, \(N=41\), \(p<0.05\)) and suicidal (332-744\,ms, two-sided cluster-based permutation test, \(N=48\), \(p<0.05\), Fig.~\ref{fig5} B, blue). Notably, \(PC_2\) showed earlier activation compared to \(PC_1\) across all groups: control (356\,ms vs.\ 332\,ms), depressed (334\,ms vs.\ 296\,ms), and suicidal (332\,ms vs.\ 280\,ms). These findings suggest that semantic processing engages a dynamic cascade of spatially and temporally organized neural representations, with distinct contributions from frontocentral (\(PC_2\)) and parietotemporal (\(PC_1\)) regions. While \(PC_2\) primarily supports early discriminative processing, \(PC_1\) may reflect more sustained integrative activity during sentence evaluation.

We observed similar temporal dynamics in the time-resolved SVM feature weights across groups; however, notable differences in their specific characteristics suggest group-specific modulation of feature contributions. To quantitatively estimate the group differences, we computed the Pearson correlation between sentence decoding accuracy and time-resolved SVM feature weights across time within each group using a bootstrap test (see Methods for details). The empirical distribution of the correlation for \(PC_1\) and \(PC_2\) revealed distinct clustering patterns for each group (Fig.~\ref{fig5} C). We found that the correlations differed significantly across all groups, showing an increasing trend in correlation from the control group to the clinical groups for both \(PC_1\) (control: 0.296, 95\% CI: -0.017--0.568, depressed: 0.624, 95\% CI: 0.39--0.79, and suicidal: 0.694, 95\% CI: 0.507--0.817, one-sample two-sided bootstrap test, \(p<0.001\), FDR-corrected, Fig.~\ref{fig5} D, left) and \(PC_2\) (control: 0.528, 95\% CI: 0.27--0.707, depressed: 0.7, 95\% CI: 0.529--0.818, and suicidal: 0.806, 95\% CI: 0.669--0.896, one-sample two-sided bootstrap test, \(p<0.001\), FDR-corrected, Fig.~\ref{fig5} D, right). Within each group, \(PC_2\) showed a significantly stronger correlation with sentence decoding compared to \(PC_1\) (control: 0.296 vs.\ 0.528, depressed: 0.624 vs.\ 0.7, and suicidal: 0.694 vs.\ 0.806, one-sample two-sided bootstrap test, \(p<0.001\), FDR-corrected). These findings suggest that the temporal dynamics and strength of the regional contributions vary systematically with mental state.

\subsection*{Cross-temporal generalization of decoding}

Cross-temporal generalization (CTG) analysis provides valuable insights into the temporal organization of information processing in the brain~\cite{king2014characterizing}. By examining patterns of the temporal generalization matrix, we can assess how neural representations are modulated and transformed over time. CTG analysis was performed by training a decoder (Fig.~\ref{fig2}C) at each timestep \(t_{train}\) and applying the resulting decision boundary across all time points \(t_{test}\). If the decision boundary estimated at one point does not generalize to the others (e.g., showing a diagonal pattern), it may indicate that the brain undergoes dynamic state transitions, reflecting temporal changes in the underlying neural representations. In contrast, if the decision boundary generalizes over multiple time points (e.g., showing a clustered pattern), it may indicate that the brain maintains a sustained representational state~\cite{king2014characterizing}.

\begin{figure*}[t]
    \centering
    \includegraphics[width=\textwidth]{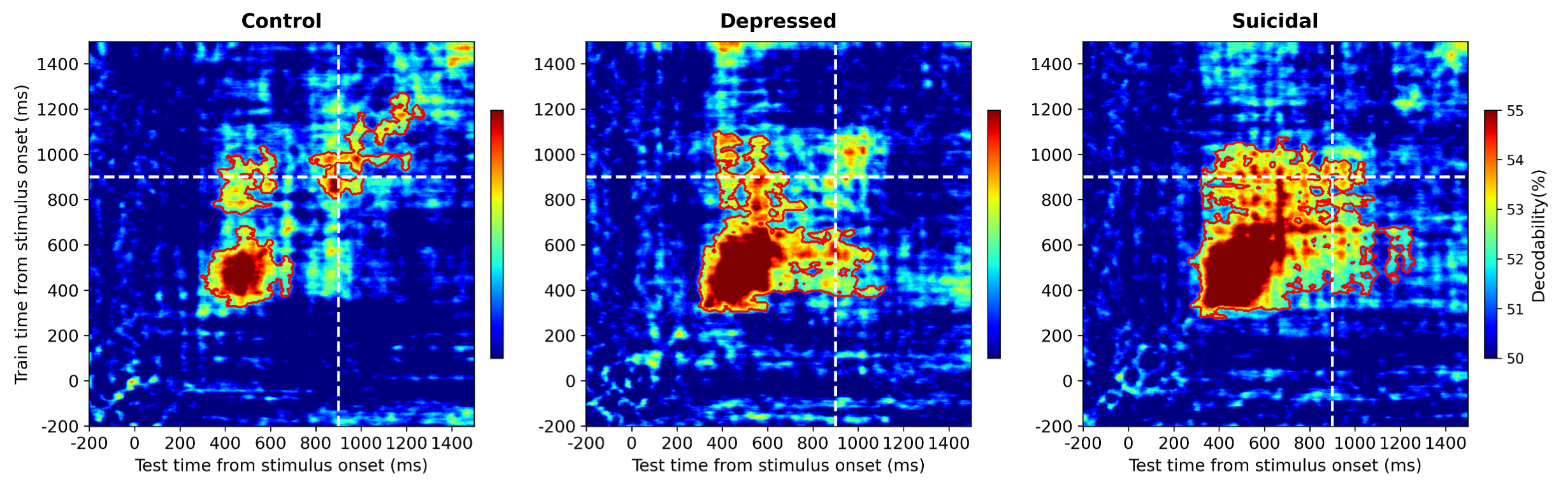}
    \caption{Cross-temporal generalization (CTG) of the sentence decoding for each group. The x-axis shows the testing time, and the y-axis shows the training time relative to the final word onset. Decoding accuracy was estimated using the multivariate EEG decoding pipeline (Fig.~\ref{fig2}) by training the classifier at each time point \(t_{train}\) and testing it at all time points \(t_{test}\), resulting in a full temporal generalization matrix. The red contour denotes clusters where decoding accuracies are significantly higher than the baseline accuracy (50\%, two-sided cluster-based permutation test, cluster-defining threshold \(p<0.05\), corrected significance level \(p<0.05\), 1000 permutations). The white dashed lines indicate the response window starting at 900\,ms.}
    \label{fig6}
\end{figure*}

Each group showed a distinct temporal generalization profile, with a progressive increase in generalization from the control to the suicidal group (Fig.~\ref{fig6}). The control group showed a strong, statistically significant diagonal cluster around 300-600\,ms, prior to the response window (after 900\,ms, two-sided cluster-based permutation test, \(N=48\), \(p<0.05\), Fig.~\ref{fig6}, left). However, no significant off-diagonal generalization was observed, suggesting temporally dynamic and modular neural processing following the initial semantic processing. In contrast to the control group, the clinical groups showed more temporally generalized patterns. The depressed group exhibited broader diagonal clusters spanning 300-700\,ms, along with sustained off-diagonal generalization extending up to 1100\,ms (two-sided cluster-based permutation test, \(N=41\), \(p<0.05\), Fig.~\ref{fig6}, middle). The suicidal group exhibited the largest and most widely generalized cluster, extending continuously from 300 to 1100\,ms (two-sided cluster-based permutation test, \(N=48\), \(p<0.05\), Fig.~\ref{fig6}, right).

Together, this progressive temporal broadening in semantic processing from control to the depressed clinical groups suggests a shift from transient, rapid, and dynamic neural responses to more sustained and temporally overlapping representations. These patterns may reflect reduced cognitive flexibility and impaired updating or disengagement from emotionally salient content, particularly among individuals with greater clinical severity~\cite{joormann2006happiness, foland2012cognitive, disner2011neural}.
\section*{Discussion}

This study examined the spatiotemporal characteristics of neural representations underlying semantic processing in healthy controls, individuals with depression without suicidality, and individuals with depression and suicidality, using EEG recordings collected during a sentence evaluation task involving emotionally charged sentences. Elucidating these neural mechanisms is critical for understanding how cognitive and affective processing diverges across varying levels of clinical severity of depression. Our results demonstrate that emotionally salient semantic content evokes distinct neural representations that systematically vary with mental health status. In a task-evoked framework, through multivariate EEG decoding, we identified a series of neural processes, spanning early semantic evaluation, sustained integration, and decision-making, and found that these dynamics were differently modulated in individuals with depression, with or without suicidality. These clinical groups exhibited decoding responses with earlier onset, prolonged duration, and greater amplitude compared to healthy controls. These differences were accompanied by altered spatial dynamics, including enhanced contributions from frontocentral and parietotemporal components, as well as broader temporal generalization patterns, suggesting sustained and overlapping neural representations in the clinical groups.

Investigating the early, task-evoked neural dynamics of semantic processing offers critical insights that extend beyond what can be inferred from behavioral responses alone. While behavioral measures can capture certain aspects of cognitive and emotional function, they can be influenced by strategies, social desirability biases, and task-related demands, and may therefore provide an incomplete or potentially misleading account of underlying cognitive and affective processes~\cite{lieberman2007social, nisbett1977telling}. Previous studies have demonstrated that semantic processing unfolds as a dynamic sequence in the brain, beginning with rapid early responses and extending into later stages associated with active evaluation and decision-making. Early EEG responses within the first 250\,ms after the stimulus onset likely reflect initial sensory encoding~\cite{hillyard_event-related_1998, lochy_robust_2015}. Shortly after, the brain engages in rapid semantic evaluation (300-600\,ms). This stage is marked by the N400 event-related potential (ERP), a well-established index of lexical-semantic integration, semantic access, and conflict monitoring when incoming content violates expectations~\cite{brouwer2013time, friederici2011brain, kutas1980reading, troyer2020catch,  lau_cortical_2008}. As comprehension unfolds, sustained activity supports the integration of meaning over time and context, extending from approximately 400 to 700\,ms~\cite{zhou_posterior_2019,binder2009semantic, valdebenito-oyarzo_parietal_2024}. Beyond 700\,ms, neural activity associated with active decision-making becomes prominent, reflecting categorical evaluation, evidence accumulation, and preparation for behavioral response~\cite{zhou_posterior_2019, gold2007neural}. Finally, conscious appraisal, response selection, and motor execution emerge in the late stage of processing, typically approaching the response period~\cite{del2007brain, dehaene2011experimental}. This temporal progression reflects how early semantic processing precedes active cognitive engagement. Based on these findings, we hypothesized that the most distinguishable neural representations for emotionally salient semantic content would emerge during the semantic evaluation and conflict-monitoring window (300-600\,ms).

In line with our expectations and with prior findings, sentence decoding revealed the most prominent differentiation of emotionally salient semantic content within the 300-600\,ms time window (Fig.~\ref{fig3}A), corresponding to the period of early semantic evaluation, integration, and conflict monitoring~\cite{brouwer2013time,friederici2011brain, kutas1980reading, troyer2020catch, lau_cortical_2008, zhou_posterior_2019, binder2009semantic}. This effect was primarily driven by frontocentral and parietotemporal components %(Supplementary Fig.~\ref{supfig2})
(Supplementary Fig.\,S2), which align with neural sources implicated in early and sustained semantic processing. The frontocentral EEG components likely reflect activity in regions such as the middle temporal gyrus (MTG), inferior frontal gyrus (IFG), and anterior cingulate cortex (ACC), which are known to support lexical-semantic access, conflict monitoring, and contextual integration~\cite{brouwer2013time, friederici2011brain, lau_cortical_2008}. These areas are reliably observed in source-localized EEG and MEG studies during the N400 time window via frontocentral electrode sites~\cite{troyer2020catch, kutas2011thirty}. In contrast, the parietotemporal components likely reflect activity in the angular gyrus and posterior MTG, which have been associated with sustained semantic integration and decision-making~\cite{zhou_posterior_2019, binder2009semantic, valdebenito-oyarzo_parietal_2024}. 

To further support our interpretation of early semantic processing, we compared these findings with the results from response decoding (Fig.~\ref{fig3}B). Robust decoding of behavioral responses emerged later, beginning at approximately 612\,ms, and was primarily driven by the parietotemporal EEG components %(Supplementary note~\ref{supfig2})
(Supplementary Fig.\,S2). This parietotemporal activity likely reflects engagement of the posterior parietal cortex (PPC), which plays a critical role in forming internal models of intention, evidence accumulation, and motor planning during decision-making~\cite{andersen_intention_2009}. This temporal dissociation suggests that neural activity related to perceptual or active decision-making and motor preparation occurred after 600\,ms, well after the peak decoding observed during sentence decoding. The non-overlapping temporal profiles of sentence and response decoding support the interpretation that neural differentiation of distinct semantic content occurs predominantly within the early window, during the 300-600\,ms. These findings further indicate that semantic evaluation, integration, and value-based differentiation are largely unaffected by active deliberation or response-related activity.

A critical question is whether the neural dynamics underlying semantic processing systematically differ across individuals with varying mental health states, particularly those experiencing depression or suicidality. Studies have demonstrated that individuals with depression show altered neural dynamics during semantic and emotional processing tasks compared to healthy controls. Reduction in N400 amplitude has been observed in frontal and central-parietal regions of depressed individuals~\cite{iakimova2009behavioral, kiang2017abnormal, klumpp2010semantic}. Sustained brain activity in response to positive or self-relevant content has been shown to be attenuated or exaggerated, depending on emotional valence, particularly in midline frontal and frontocentral regions~\cite{shestyuk2005reduced, benau2019increased}. fMRI and EEG studies further suggest that the PPC, ACC, and default mode network (DMN) exhibit altered functional connectivity and engagement in depression~\cite{gallo2023functional, pilmeyer2022functional, koo2017current}. Moreover, prolonged neural activity during semantic and emotional processing tasks has been observed in clinical populations, suggesting deficits in cognitive flexibility, heightened attentional bias, and difficulties disengaging from emotionally salient content~\cite{joormann2006happiness, foland2012cognitive, disner2011neural}. Collectively, these studies support the interpretation that the spatiotemporal neural dynamics of semantic evaluation and self-referential processing are systematically altered in clinical populations. Our results provide empirical support for this interpretation. Sentence decoding revealed differences between control, depressed, and suicidal individuals, specifically within the early semantic processing window (300-600\,ms, Fig.~\ref{fig4}), consistent with the time frame of lexical-semantic evaluation and conflict monitoring~\cite{brouwer2013time, friederici2011brain, kutas1980reading}. The control group exhibited a relatively sharp, temporally confined decoding peak, aligned with rapid and efficient semantic evaluation as reported in typical N400 dynamics~\cite{troyer2020catch, kutas2011thirty}. In contrast, both depressed and suicidal groups showed earlier onset, prolonged duration, and enhanced decoding responses, indicating heightened and sustained engagement with emotionally salient material. This may reflect the altered sensitivity to negative or self-referential stimuli in depression and suicidality~\cite{shestyuk2005reduced, benau2019increased}. 

These effects were primarily driven by enhanced engagement in both frontocentral and parietofrontal EEG components (Fig.~\ref{fig5}), corresponding to neural sources involved in conflict monitoring, semantic integration, and value-based evaluation~\cite{zhou_posterior_2019, binder2009semantic, lau_cortical_2008}. The joint contribution and alteration of these components may reflect altered dynamics within the frontoparietal network (FPN), which supports cognitive control and flexible semantic processing, and the DMN, which has been implicated in self-referential processing and emotional evaluation~\cite{seeley_dissociable_2007, menon_large-scale_2011, andrewshanna_default_2014}. The broader and more sustained temporal generalization profiles observed in the clinical groups (Fig.~\ref{fig6}) may further support the presence of impaired cognitive flexibility and difficulties in disengaging from emotionally salient or self-referential content. These spatiotemporal alterations suggest a shift from efficient, transient semantic encoding in healthy controls to prolonged, overlapping, and potentially maladaptive neural representations in those with greater clinical severity~\cite{joormann2006happiness, foland2012cognitive, disner2011neural}. 

In summary, our findings advance the understanding of how emotionally salient semantic content is processed in the brain and how these neural processes differ across individuals with different mental health states. By leveraging multivariate decoding of EEG data, we were able to detect fine-grained, dynamic neural representations of semantic processing at a whole-brain scale, offering spatiotemporal interpretability beyond what traditional univariate approaches afford~\cite{cichy_multivariate_2017, marsicano_decoding_2024}. The use of an active sentence evaluation task, in which subjects explicitly engaged with emotionally charged statements, further ensured that the observed neural dynamics reflected meaningful semantic affective integration, rather than passive language perception. Moreover, the identified spatiotemporal characteristics, spanning early semantic evaluation, sustained integration, and delayed decision-related activity in clinical groups, may provide a principled foundation for extracting meaningful neural features, potentially guiding the development of more precise and interpretable classification models for assessing individual mental health~\cite{simmatis_technical_2023}. Future research integrating source localization~\cite{leahy_study_1998,baillet_electromagnetic_2001} and time-frequency analysis may offer a more comprehensive understanding of the underlying neural mechanisms. Incorporating these spatiotemporal neural signatures into deep learning frameworks holds potential to enhance early detection and enable more precise assessment of mental health conditions.
\section*{Methods}

\subsection*{Participants}

160 participants (99 females) between 18 and 25 years of age from the University of Southern California (USC) were recruited for this experiment. All experimental protocols were approved by the Institutional Review Board (IRB) of USC (UP-23-00071). Participants were informed about the experimental contents before the sessions. All participants were fluent in English and were asked to complete a set of questionnaires to determine their eligibility for the study. These questionnaires included the Patient Health Questionnaire-9 (PHQ-9)~\cite{kroenke2001phq} and the Suicidal Ideation Scale (SIS)~\cite{rudd1989prevalence}, measuring their stress, depression, and risk of suicide. During recruitment, participants were pre-screened into three groups based on their PHQ-9 and SIS scores: healthy control (C, PHQ-9 $\leq4$), depressed non-suicidal (D, PHQ-9 $\geq10$ and SIS $\leq16$), and depressed suicidal (S, PHQ-9 $\geq10$ and SIS $>16$). Individuals were excluded if they met any of the following criteria: 1) current or previous diagnosis of neurological and psychiatric disorders, including Schizophrenia, Bipolar Disorder, Epilepsy, Brain Cancer, and Stroke, 2) color blindness, 3) having learned English after the age of 7, and 4) a PHQ-9 score between 5 and 9. Eligible participants completed additional questionnaires on the day of the EEG recording, including the PHQ-9, the Generalized Anxiety Disorder 7-item scale (GAD-7)~\cite{williams2014gad}, and the Ten-Item Personality Inventory (TIPI)~\cite{gosling2003very}. Out of 160 participants, we excluded 13 participants with unstable PHQ-9 scores between the day of the recruitment and the day of the experiment (i.e., if their score changed by more than 5 points, resulted in reclassification into a different group, or placed them within the exclusion range). We further discarded 10 participants with bad EEG recordings by visually inspecting the data during the EEG preprocessing. From the remaining 137 participants, there were 48 participants in Group C, 41 in Group D, and 48 in Group S.

\subsection*{Experimental design and stimuli}
To investigate the spatiotemporal dynamics of the neurocognitive process, participants performed a visual sentence evaluation task (Fig.~\ref{fig1}A). A total of 160 sentences with 4 to 14 words were designed to elicit neural responses depending on whether the sentence is congruent or incongruent with the participant's beliefs, prior knowledge, experiences, or values. We constructed 80 distinct sentence pairs, each identical except for the final (critical) word, which differed between the two sentences. In each pair, the final words were selected such that one conveyed semantics likely congruent for the participant while the other conveyed semantics likely incongruent for the participant. In this study, congruent sentences refer to statements that suicidal participants were likely to agree with. In contrast, incongruent sentences refer to those that suicidal participants were likely to disagree with; (e.g., congruent: ``My mind gravitates toward thoughts of sadness" versus incongruent: ``My mind gravitates toward thoughts of joy"). Each sentence pair belonged to one of four topics of interest (TOIs): (1) neutral biographical information, (2) depression-related actions, (3) depression-related declarative reflection, and (4) intentions related to suicidal ideation, with 20 sentence pairs in each TOI.

Each trial started with a fixation cross presented at the center of the screen for 300 ms. The sentences were presented word by word, with each word presented for 300 ms with a 300 ms inter-stimulus interval (ISI) until the final word. The final word was presented for 600 ms, followed by a 300 ms ISI with a black screen, and followed by a 2-s response window. During the response time, the words ``agree" and ``disagree" were presented on the screen. Participants were asked to respond whether they agreed or disagreed with the statement by pressing a button with their dominant hand. The experiment comprised 4 blocks, and each block consisted of 80 trials with 40 congruent and 40 incongruent sentences. The full set of 160 sentences was randomly presented in blocks 1 {\&} 2 and repeated in blocks 3 {\&} 4 with randomization. The recording session lasted about 55 minutes. A detailed description of the experimental design can be found in~\cite{hughes2025precog}.

\subsection*{EEG recording and preprocessing}
We collected EEG data using a 64-channel active electrode system (actiCAP, Brain Products, GmbH) with a 1~kHz sampling rate. The impedances of the electrodes were initially lowered to 5 k$\Omega$ by injecting an electrolyte gel (SuperVisc Gel, Brain Products, GmbH) before the recording and were maintained under 30 k$\Omega$ throughout the EEG recording. We built a semi-automated EEG preprocessing pipeline in MATLAB (Mathworks, Inc.) primarily using Brainstorm~\cite{tadel2011brainstorm} together with EEGLab~\cite{delorme2004eeglab} and FieldTrip toolbox~\cite{oostenveld2011fieldtrip}. We removed the powerline noise with a 60 Hz IIR notch filter and filtered the data from 0.5--80 Hz using a FIR band-pass filter in Brainstorm. Bad channels and bad time segments were detected using the Artifact Subspace Reconstruction (ASR) routine~\cite{mullen2015real} and the clean{\_}rawdata function in EEGLab with the following criteria: a) channels with flatline longer than 4 seconds, b) correlation with the neighboring channels less than 0.8, and c) ASR threshold of 50~\cite{mullen2015real}. The labeled bad segments were imported into Brainstorm for subsequent pre-processing. Bad channels were removed and then interpolated by averaging the neighboring channels within a 5~cm distance. Bad time segments were temporarily removed from the data before Independent Component Analysis (ICA)~\cite{lee1998independent} for artifact removal. Following this, we performed ICA and manually removed artifactual components (eye blinks, eye movements, electrocardiogram, and muscle movements). The remaining clean components were used to reconstruct the signal across the entire dataset. We applied the same bad time-segments detection routine to this cleaned data to capture the remaining bad time segments after the artifact removal. This approach allowed us to retain a greater portion of the data compared to the initial bad segment detection, minimizing the data loss. The preprocessed data were re-referenced to the average of all channels, and the linear trend was removed. Finally, we epoched the trials from -200 ms to 1500 ms from the stimulus onset and z-scored every trial using the time window between -200 ms and stimulus onset. In this study, we excluded 17 channels (Fp1, Fp2, AF7, AF8, F7, F8, FT7, FT8, FT9, FT10, T7, T8, TP7, TP8, TP9, TP10, and Iz) in the outer boundary that were prone to muscle activity and noise, and only used the remaining 47 channels (Fig.~\ref{fig2}A). The EEG data were downsampled to 250 Hz and low-pass filtered to 20 Hz using a zero-phase Butterworth IIR low-pass filter~\cite{oostenveld2011fieldtrip, tadel2019meg}.

\subsection*{EEG multivariate pattern analysis}
We used multivariate pattern analysis to investigate the spatiotemporal information of the neural response within each subject (Fig.~\ref{fig2}). We used a stratified 5-fold cross-validated~\cite{berrar2019cross} pairwise classification accuracy of the linear support vector machine (SVM)~\cite{hearst1998support, dobs2019face} to measure the EEG pattern dissimilarity between pairs of stimuli. In each subject, trials were divided into five stratified folds, ensuring that each fold contained approximately the same proportion of trials from each condition (congruent vs.\ incongruent). In each of the five cross-validation iterations, one fold was held out as the test set while the remaining four folds were used for training. This process was repeated such that each fold served as the test set exactly once, ensuring that all trials were used for both training and testing without repetition within any single iteration. To extract the latent factors shared across groups, we first generated average group responses by averaging trials in the training set in each subject and subsequently averaging them across subjects in each group. This procedure was designed to emphasize consistent, group-level spatial patterns of neural activity while minimizing inter-subject variability. The resulting group-averaged data were then concatenated across time and used to identify latent components that capture spatial features common across groups. We applied principal component analysis (PCA)~\cite{abdi2010principal} across spatial dimensions (i.e., channels) and extracted the weights $(\bm{W} \in \mathbb{R}^{3 \times 47})$ corresponding to the first 3 principal components (PCs) with $>$ 95\% cumulative explainable variance (Fig.~\ref{fig2}B). $\bm{W}$ was used as a projection operator into the latent feature space.

The full EEG data for each subject, encompassing both training and test sets, $\bm{E} \in \mathbb{R}^{47 \times time \times trial}$, was projected onto a low-dimensional latent space by the projection matrix $\bm{W}$:
\begin{equation}
    \bm{\hat{E}} = \bm{W E} \quad \text{where } \bm{\hat{E}} \in \mathbb{R}^{3 \times time \times trial}
\end{equation}

To enhance the signal-to-noise ratio (SNR) and facilitate robust decoding, we augmented trials using a bootstrapping and sub-averaging procedure~\cite{murphy2022decoding}. We augmented 400 train trials (200 congruent and 200 incongruent) and 100 test trials (50 congruent and 50 incongruent). Each augmented trial was generated separately for the training and test sets by randomly sampling 12 trials (approximately 10 \% of the trials in each congruency condition of the training set) with replacement within each congruency condition and training/testing cohort and computing the average across samples. This approach ensured strict separation between training and testing sets, thereby preventing data leakage and preserving the integrity of the decoding analysis. At each time $t$, we performed binary classification between congruent and incongruent bootstrapped trials. This PCA-based feature extraction, bootstrapping, and classification procedure was repeated across each cross-validation fold, and the average accuracy across the five folds was reported as the measure of decodability (\%) for each subject (Fig.~\ref{fig2}C). The alternative analysis was done separately, from a training/testing split through classification, by relabeling trials based on the subject's response to the sentence (agree vs.\ disagree)\textemdash regardless of the congruency\textemdash to probe the spatiotemporal dynamics of decision-related neural processes. 

To examine how the temporal dynamics of neural representations generalize across time, we applied cross-temporal generalization analysis~\cite{king2014characterizing}. The classifier was trained on the data at training time $t_{train}$ and tested on the data at test time $t_{test}$. This procedure was repeated for all pairs of $t_{train}$ and $t_{test}$.

\subsection*{Assessing group differences in spatiotemporal dynamics}
To examine group differences in the temporal dynamics of decoding, we assessed statistical differences in onset latency (i.e., the earliest significant time point of the significant cluster containing the peak decoding accuracy), offset latency (i.e., the latest significant time point of the significant cluster containing the peak decoding accuracy), peak latency (i.e., the time point of maximum decoding accuracy), and peak amplitude (i.e., accuracy at the peak latency) between subject groups using bootstrap tests. Empirical null distributions for each latency and amplitude measure were inferred by randomly bootstrapping subject-specific decoding time series within each group for 1000 repetitions. The \(2.5^{th}\) and the \(97.5^{th}\) percentiles of the distribution defined the 95\% confidence intervals (CIs). 

To assess group differences in the spatial dynamics of decoding, we computed the Pearson correlation between sentence decoding accuracy and the time-resolved SVM feature weight across the decoding window using bootstrap tests. Empirical null distributions for each decoding accuracy and SVM feature weight were derived by averaging randomly bootstrapped subject-specific time series within each group for 1000 repetitions. The correlation was computed for each pair of decoding accuracy and the time-resolved absolute SVM weights of \(PC_1\) and \(PC_2\) in the 1000 bootstrapped samples. The \(2.5^{th}\) and the \(97.5^{th}\) percentiles of the distribution defined the 95\% confidence intervals (CIs).

To test the statistical significance between the groups, we derived the null distribution by randomly permuting samples between groups 1000 times. The \(p\)-value of the difference between groups was defined as the number of absolute differences in the null distribution that exceed or equal to the absolute difference of the observation, divided by the number of permutations (two-sided permutation test). We applied the false discovery rate (FDR) correction using the Benjamini-Hochberg procedure~\cite{benjamini1995controlling} to compensate for the multiple comparison problem.

\subsection*{Statistics}
We performed non-parametric cluster-based permutation \textit{t}-test~\cite{maris2007nonparametric, pantazis2005comparison} for all time-series analyses to determine whether the observation is significantly different from the null hypothesis. The null hypothesis assumes a 50\% chance level for decoding accuracies, and 0 for the weight estimation. Significant temporal clusters were defined from the initial test statistics by selecting consecutive significant time stamps. Then we randomly permuted the samples between the observation and the null hypothesis at each time point within the clusters, and ran test statistics. This procedure was repeated 5000 times to construct the permutation distribution. If the observed test statistics exceed the \(95^{th}\) percentiles of the permutation distribution, it was considered significant (\(p\)$<$0.05, one-sided). The corrected \(p\)-value was computed by calculating the number of test statistics that were larger than the observed test statistics.

\section*{Data Availability}
Anonymized EEG recordings and annotations used in this study are available upon request to shri@usc.edu.

\section*{Code Availability}
The code used in this study is available upon request at woojaeje@usc.edu. EEG preprocessing was performed in MATLAB (R2024a) using Brainstorm (\url{https://neuroimage.usc.edu/brainstorm/}), EEGLAB (\url{https://sccn.ucsd.edu/eeglab/}), and the FieldTrip toolbox (\url{https://www.fieldtriptoolbox.org/}). EEG decoding analyses were performed in Python 3.12 using the Scikit-learn library. 

\bibliography{references}

\section*{Acknowledgments}
This study was sponsored by the Defense Advanced Research Projects Agency (N660012324006) and by the National Institute of Biomedical Imaging and Bioengineering (R01EB026299). The content of the information does not necessarily reflect the position or the policy of the Government, and no official endorsement should be inferred.

\section*{Funding}
This study was supported by DARPA under N660012324006 and NIBIB under R01EB026299.

\section*{Author Contributions}
W.J. preprocessed the EEG data, analyzed the behavioral and EEG data, performed computational modeling, and wrote the initial draft of the manuscript.
W.J., T.M., D.P., and R.L. conceived the EEG decoding method.
D.B., A.H., R.C., I.B., K.L., S.N., and R.L. conceptualized the study.
D.B., A.H., R.C., I.B., and Th.M. designed the experimental protocol.
W.J., T.M., M.H., and C.M. implemented the experimental protocol and acquired behavioral and EEG data.
W.J., A.K., K.A., C.M., E.K., D.B., A.H., R.C., D.P., S.K., T.M., S.N., and R.L. contributed to the interpretation of the results.
All authors contributed to the revision and editing of the manuscript.

\section*{Competing Interests}
The authors declare no competing interests.

\newpage

\section*{Supplementary Notes}
\renewcommand{\thefigure}{S\arabic{figure}}
\setcounter{figure}{0} % if this is your first supplementary

\begin{figure}[H]
    \centering
    \includegraphics[width=0.6\textwidth]{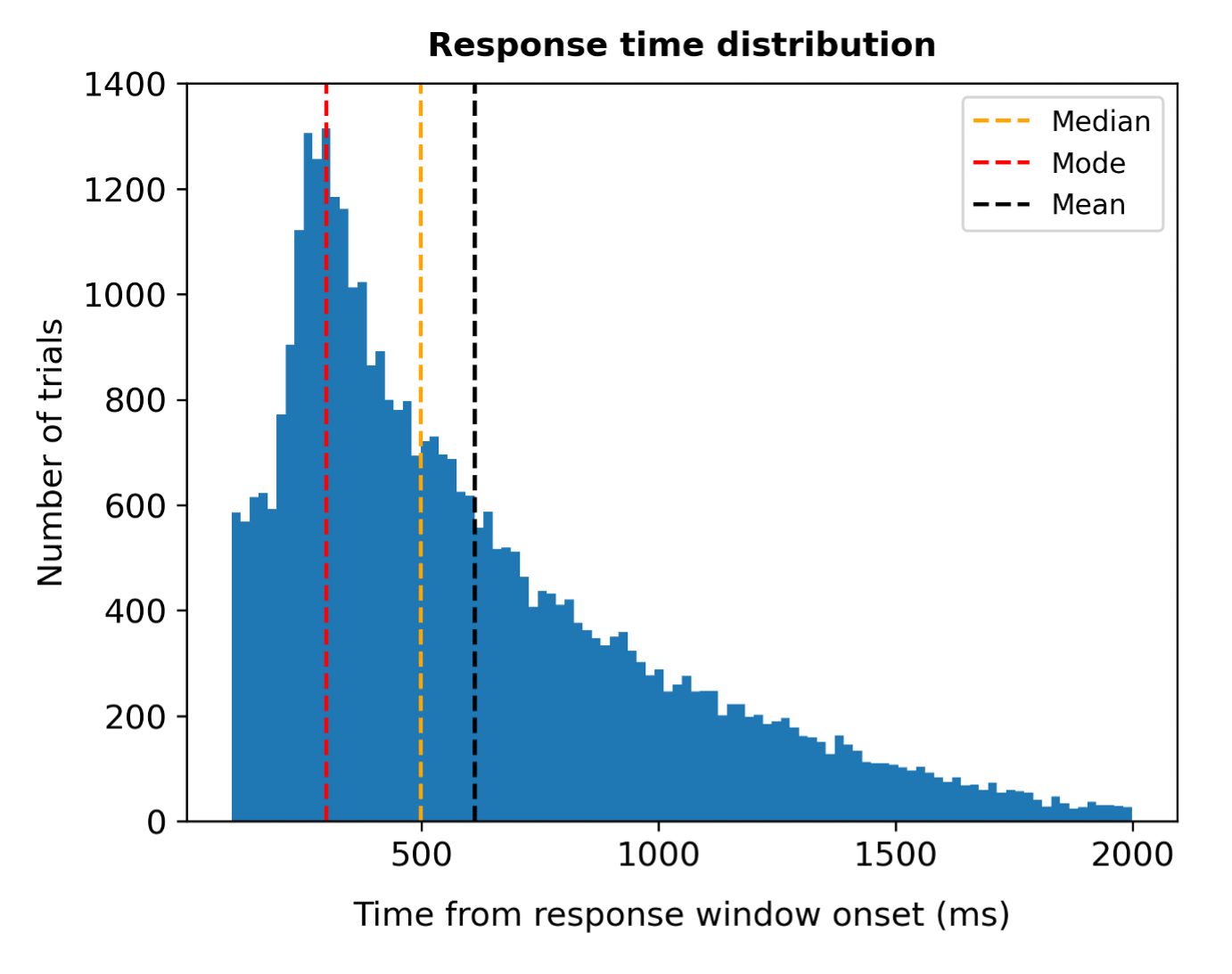}
    \caption{Distribution of response time across all trials, measured from the onset of the response window. Trials with less than 100 ms response time were excluded. The black dashed line indicates the mean response time (613 ms), the orange dashed line marks the median (499 ms), and the red dashed line denotes the mode (299 ms).}
    \label{supfig1}
\end{figure}

\newpage
\renewcommand{\thetable}{S\arabic{table}}

\begin{table}[ht]
\centering
\begin{tabular}{lll}

\toprule
\textbf{Sentence type} & \textbf{Sentiment} & \textbf{Counts}
\\
\midrule

\multirow{3}{*}{Congruent} 
    & Positive & 2
    \\
    & Negative & 62
    \\
    & Neutral & 16
    \\
\midrule

\multirow{3}{*}{Incongruent} 
    & Positive & 42
    \\
    & Negative & 7
    \\
    & Neutral & 31
    \\
\bottomrule
\end{tabular}
\caption{Sentences were labeled as congruent or incongruent based on the expected response of the suicidal group to the sentence. To characterize the emotional content of each sentence type, we counted the number of sentences with positive, neutral, and negative sentiment within the congruent and incongruent categories.}
\label{tabs1}
\end{table}

\newpage
\renewcommand{\thetable}{S\arabic{table}}

\begin{table}[ht]
\centering
\begin{tabular}{llll}

\toprule
\textbf{Group} & \textbf{Sentence type} & \textbf{Response type} & \textbf{Counts (mean \(\boldsymbol{\pm}\) SD)}
\\
\midrule

\multirow{4}{*}{Control} 
    & \multirow{2}{*}{Congruent} & Agree & \(51.9\pm13.18\)
    \\
    &                            & Disagree & \(105\pm14.24\)
    \\
    & \multirow{2}{*}{Incongruent} & Agree & \(103.42\pm12.14\)
    \\
    &                            & Disagree & \(52.13\pm11.17\)
    \\
\midrule

\multirow{4}{*}{Depressed}
    & \multirow{2}{*}{Congruent} & Agree & \(81.73\pm18.05\)
    \\
    &                            & Disagree & \(74.2\pm18.21\)
    \\
    & \multirow{2}{*}{Incongruent} & Agree & \(78.2\pm19.22\)
    \\
    &                            & Disagree & \(77.54\pm17.37\)
    \\
\midrule

\multirow{4}{*}{Suicidal}
    & \multirow{2}{*}{Congruent} & Agree & \(107.81\pm17.85\)
    \\
    &                            & Disagree & \(49.33\pm16.72\)
    \\
    & \multirow{2}{*}{Incongruent} & Agree & \(63.4\pm14.79\)
    \\
    &                            & Disagree & \(93.65\pm15.86\)
    \\
\bottomrule
\end{tabular}
\caption{Mean response counts (\(\pm\) standard deviation) for each group (Control, Depressed, and Suicidal) by sentence type (Congruent vs. incongruent) and response type (Agree vs. Disagree).}
\label{tabs2}
\end{table}

\newpage
\begin{figure}[H]
    \centering
    \includegraphics[width=0.9\textwidth]{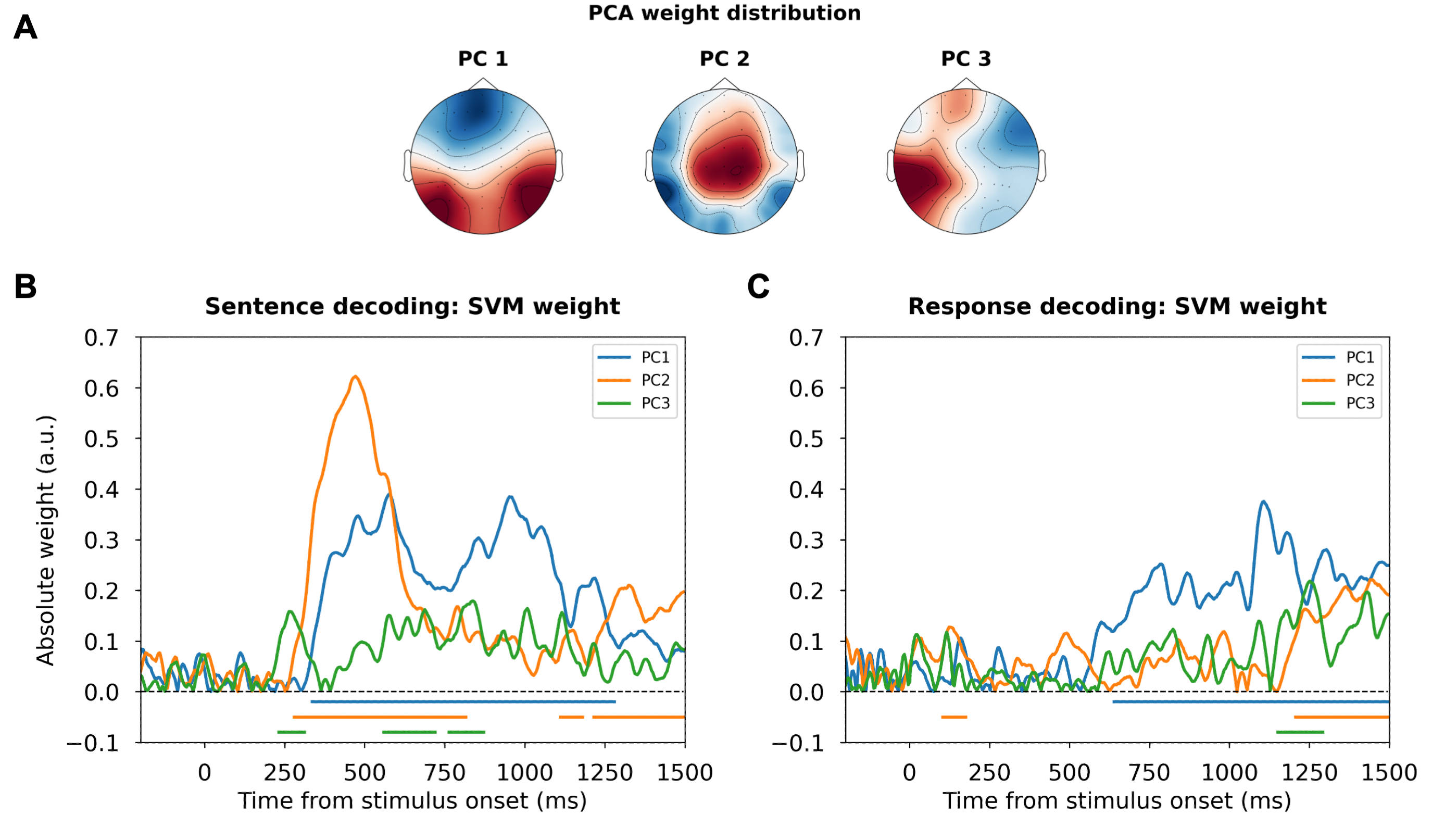}
    \caption{\textbf{Spatial contribution estimates}
    \textbf{A.} Topographical distributions of the top 3 PCs estimated from the multivariate analysis (Fig.~\ref{fig2}B). Red-highlighted regions indicate the enhanced channel contribution, and the blue-highlighted regions indicate the suppressed channel contribution.
    \textbf{B, C.} Mean estimation of the weight assigned to each of the 3 PCs by the linear SVM across subjects at each time point relative to the target word onset for sentence decoding (\textbf{B}) and response decoding (\textbf{C}). The colored lines at the bottom of the plot denote the significant time points (two-sided cluster-based permutation test, \(N=137\), cluster-defining threshold \(p<0.05\), corrected significance level \(p<0.05\), 5000 permutations).}
    \label{supfig2}
\end{figure}

\(PC_1\) showed significant contribution from 336 to 1280 ms (two-sided cluster-based permutation test, \(N=137\), \(p<0.05\), Fig.~\ref{supfig2}B, blue), corresponding to the time window where we found significant sentence decodability (304-1036 ms, Fig.~\ref{fig3}A). \(PC_2\) exhibited two temporally distinct significant clusters: an early cluster from 280 to 816 ms, and the later cluster around the response period (1112-1180 ms and 1216-1500 ms, two-sided cluster-based permutation test, \(N=137\), \(p<0.05\), Fig.~\ref{supfig2}B, orange). These clusters respectively overlap with the window associated with semantic perception and decision-making, and with the response preparation and execution window. \(PC_3\) did not show consistent significant decoding weights (Fig.~\ref{supfig2}B, green); however, the early cluster (232-312 ms, cluster-based permutation test, \(N=137\), \(p<0.05\)) may reflect the early language-dominant semantic processing and the brief engagement in the later clusters (560-720 ms and 764-872, cluster-based permutation test, \(N=137\), \(p<0.05\)) potentially reflecting transient involvement in context-sensitive semantic processing. 

In the response decoding, \(PC_1\) exhibited a significant cluster from 636 to 1492 ms (cluster-based permutation test, \(N=137\), \(p<0.05\), Fig.~\ref{supfig2}C, blue) overlapping with the period of significant response decodability (612-1496 ms, Fig.~\ref{fig3}B). Throughout this interval, \(PC_1\) showed prominent involvement until the response execution window (600-1200 ms, Fig.~\ref{supfig2}C, blue), reinforcing its role in sustained semantic integration and decision-related processing. \(PC_2\) became significantly engaged during the response execution window (1204-1492 ms, cluster-based permutation test, \(N=137\), \(p<0.05\), Fig.~\ref{supfig2}C, orange), consistent with its late re-engagement observed in the sentence decoding results.

Together, these findings offer a systematic characterization of the spatiotemporal dynamics underlying neural representations that most effectively differentiate sentence types and responses. Sentences with distinct semantic contents were most effectively discriminated during the early time window (300-600 ms), primarily within the fronto-central regions (\(PC_2\)) associated with evaluative and conflict-monitoring processes~\cite{friederici2011brain}, alongside parietotemporal regions (\(PC_1\)) involved in the semantic integration process~\cite{brouwer2013time}. Different response types were most effectively differentiated in the later window (600-1200 ms), with dominant contributions from parietotemporal regions (\(PC_1\)), reflecting sustained semantic integration processes and decision-making~\cite{troyer2020catch, valdebenito-oyarzo_parietal_2024, andersen_intention_2009}.
\end{document}